\documentclass[preprint,12pt]{elsarticle}

 \usepackage{graphicx}
 \usepackage{epsfig}

\usepackage{amssymb}

\journal{Astroparticle Physics}

\begin{document}

\begin{frontmatter}



\title{Radiopurity of Micromegas readout planes}


\author[UNIZAR]{S. Cebri\'{a}n}\author[UNIZAR]{T. Dafni}\author[SACLAY]{E. Ferrer-Ribas}\author[UNIZAR]{J. Gal\'{a}n}\author[SACLAY]{I. Giomataris}\author[UNIZAR]{H. G\'{o}mez\corref{cor1}}\ead{hgomez@unizar.es}\author[UNIZAR]{F.J. Iguaz\fnref{add1}}\author[UNIZAR]{I.G. Irastorza}\author[UNIZAR]{G. Luz\'{o}n}\author[CERN]{R. de Oliveira}\author[UNIZAR]{A. Rodr\'{i}guez}\author[UNIZAR]{L. Segu\'{i}}\author[UNIZAR]{A. Tom\'{a}s}\author[UNIZAR]{J.A. Villar}

\address[UNIZAR]{Laboratorio de F\'{i}sica Nuclear y Astropart\'{i}culas, Universidad de Zaragoza, 50009 Zaragoza, Spain}
\address[SACLAY]{CEA, IRFU, Centre d'etudes de Saclay, 91191 Gif-sur-Yvette, France}
\address[CERN]{European Organization for Nuclear Research (CERN), CH-1211 Gen\`{e}ve, Switzerland}

\cortext[cor1]{Corresponding Author: Laboratorio de F\'{i}sica Nuclear y Astropart\'{i}culas, Facultad de Ciencias, Pedro Cerbuna 12, 50009 Zaragoza, Spain. Phone number: +34 976761246. Fax number: +34 976761247}

\fntext[add1]{Present address: CEA, IRFU, Centre d'etudes de Saclay, 91191 Gif-sur-Yvette, France}

\begin{abstract}
Micromesh Gas Amplification Structures (Micromegas) are being used in an increasing number of Particle Physics applications since their conception fourteen years ago. More recently, they are being used or considered as readout of Time Projection Chambers (TPCs) in the field of Rare Event searches (dealing with dark matter, axions or double beta decay). In these experiments, the radiopurity of the detector components and surrounding materials is measured and finely controlled in order to keep the experimental background as low as possible. In the present paper, the first measurement of the radiopurity of Micromegas planes obtained by high purity germanium spectrometry in the low background facilities of the Canfranc Underground Laboratory (LSC) is presented. The obtained results prove that Micromegas readouts of the \textit{microbulk} type are currently manufactured with radiopurity levels below 30 \textrm{$\mu$Bq/cm$^{2}$} for Th and U chains and $\sim$60 \textrm{$\mu$Bq/cm$^{2}$} for $^{40}$K, already comparable to the cleanest detector components of the most stringent low background experiments at present. Taking into account that the studied readouts were manufactured without any specific control of the radiopurity, it should be possible to improve these levels after dedicated development.
\end{abstract}


\end{frontmatter}


\section{Introduction}

The quest for the direct detection of dark matter in the form of Weakly Interacting Massive Particles (WIMPs) \cite{Irastorza}, the search for the neutrinoless double beta decay ($0\nu\beta\beta$) \cite{Avignone} or other experiments involving very low rate processes (the so-called Rare Event searches) impose very challenging experimental requirements. The most important of these requirements is the reduction of experimental backgrounds down to ever-decreasing levels. The fight against backgrounds is performed in several ways, including active and passive shielding, operation in underground sites, event discrimination and especially selecting carefully the materials composing the detector and those in the surroundings so that they have the lowest radioactive traces that could induce background in the experiment. All these practices have evolved toward a technical field itself, usually called \emph{ultra-low background techniques} \cite{Heusser, Formaggio}.

It is nowadays an adopted practice within low background experiments to exhaustively screen all materials and components present in the inner part of the experimental set-up. Several detection techniques have been developed and used for low level counting in the last three decades like for example proportional counters or NaI(Tl) detectors. Nevertheless, $\gamma$-spectrometry using germanium detectors is, in most of the cases, the best way to make radiopurity measurements in a non-destructive way due to good intrinsic features of these detectors like their energy resolution. Gamma spectrometry, together with neutron activation analysis or mass spectrometry, are at present the most commonly used techniques in order to determine the radioactivity level of different materials.

For the latest generation of Rare Event experiments, the requirements for radiopurity are so strong that this constitutes an important selection criteria for the detection technique itself. Compact and simple detector set-ups are preferred as they are supposed to be better subjected to radiopurity validation, i.e. scrupulous screening of all detector components, identification of potential
sources of radioactivity and development of substitute radiopure components if possible. Complex detectors are, a priori, more likely to contain radioactive materials in their fabrication and it is therefore more difficult (or impossible) to develop radiopure versions of them. This argument gets stronger when larger detector masses and volumes are needed, as is the case of the new generation of Rare Events experiments.

Although gaseous Time Projection Chambers (TPCs)could offer unique features to Rare Event detection (as it will be described in Section \ref{RES&TPC}), they have been historically considered not suitable for these searches. The reasons for this are mainly the complexity that the construction of readouts based on one or several plane of wires (Multi Wire Proportional Chambers, MWPC) implies when it comes to large volumes of detection, mechanical and electronics-wise; but also issues related to the large volumes (e.g. difficulty to shield) and other features of the detection in gas (e.g. stability and homogeneity issues) which are not best suited for very long exposure high sensitivity operation.

However, this view is changing in the last years, thanks to several developments on TPC readouts, like the ones regarding Micromegas (Micromesh Gas Structure) readout planes, briefly explained in Section \ref{MM_readout_plane}, that are overcoming the traditional limitation of gas TPCs for Rare Event searches. Indeed, several ongoing initiatives plan to use TPCs, and in particular Micromegas readouts, in the field of Rare Event searches as briefly explained in section \ref{RES&TPC}. The present work is focused on a point of common interest for all these experiments: the radiopurity of Micromegas readouts themselves. The ultimate goal of this line of development is not only to assess the radiopurity of every type of current Micromegas readout (that has been partially studied and is presented in this article), but also to identify the less radiopure components (whether raw materials or fabrication processes introducing radioactive traces), find alternative, more radiopure ones and to implement them in the manufacturing processes.

In this article first results of measurements of Micromegas samples performed in the Canfranc Underground Laboratory (LSC) with an ultralow background high purity germanium detector are presented, including a first assessment on the state-of-the-art of Micromegas fabrication processes from the point of view of potential radiopurity. In section \ref{RES&TPC}, a summarized review of utilization of TPCs for Rare Event searches and the improvements achieved is presented. Section \ref{MM_readout_plane} presents a description of the Micromegas readouts and some details on the fabrication materials and processes that may be of relevance for the goal are given. The experimental set-up used to carry out the measurements as well as the description of the samples measured is described in section \ref{setup}, while the obtained results are presented in section \ref{RandD}. To finish, preliminary conclusions and outlook are developed in section \ref{Conclusions}.

\section{Gas TPCs for Rare Event searches}\label{RES&TPC}

It has long been recognized that gas TPCs could offer unique features to Rare Event detection. Gas TPCs provide very useful topological information that allows better identification of Rare Events discriminating them from background events. In the field of Dark Matter WIMP detection, although tests are ongoing with directional-sensitive solid detectors or nuclear emulsions \cite{Ahlen}, gas TPCs present the most promising technique to measure the direction of the WIMP-induced nuclear recoils. After a WIMP collision, the nuclei recoil with typical energies of 1-100 \textrm{keV}, travel distances of the order of few 100 \textrm{{\AA}} in solids, while in gases this distance can go up to the \textrm{mm} or \textrm{cm} scale, depending on the type and pressure of the gas. An accurate measurement of the recoil direction would provide an unmistakable signature of a WIMP, which no terrestrial background could mimic \cite{Spergel}. In the decade of the 1990s, only one group, the DRIFT collaboration \cite{Pipe}, was exploring the use of MWPC-read TPCs for this purpose. Despite very recent encouraging results, the accurate measurement of the direction of low energy nuclear recoils with MWPCs remains very challenging. Due to the mechanical limitations in the construction of the planes of wires, the best readout granularity achievable by these devices lies in the \textrm{mm} scale, very limiting, if not insufficient, for an accurate imaging of nuclear recoil tracks. Only the advent of micropattern readouts and Micromegas-like structures, has recently allowed more groups to  further explore the detection of WIMP directionality with gas TPCs \cite{Ahlen,Dujmic}.

In the search for the $0\nu\beta\beta$ decay a similar situation can be drawn. The topology of the $0\nu\beta\beta$ signal in a gas TPC with a proper 3D readout (two electron tracks ending in two ionization blobs) is a powerful tool for background event rejection and signal identification. This fact was proved in the 1990s by the Gothard xenon gaseous TPC, which aimed to measure the $0\nu\beta\beta$ signal of the $^{136}$Xe \cite{Luscher}. This experiment was an important milestone, successfully demonstrating the proof-of-principle. However, the technique was not further pursued because of the drawbacks of MWPCs and TPCs mentioned above, and the fact that the obtained energy resolution of MWPCs in Xe was insufficient for the next generation of double beta decay experiments.

However, in recent years important technical advantages in the readouts of TPCs have been achieved that may eliminate the traditional limitations of these detectors, opening new windows of application for them. In particular, Rare Event searches could now profit enormously from the signal identification potentiality of gas TPCs. The key issue that has boosted the development
activities of these devices has been the use of metallic strips or pads, precisely printed on plastic supports with photolithography techniques (much like printed circuit boards). These strips substitute the traditional wires to receive the drifting charge produced in the gas. The simplicity, robustness and mechanical precision are much higher than those of conventional planes of wires. Around this basic principle, first introduced by Oed \cite{Oed} already in 1988 under the name Microstrip Gaseous Chamber (MSGC), several different designs have been developed, which differ in the way the multiplication structure is done, and that in general are referred to as Micro Pattern Gas Detectors (MPGD). One of the most promising MPGDs, and certainly the one offering most attractive features for application in Rare Events, is the Micromegas readout plane \cite{Giomataris_I}.

The Micromegas concept dates back to 1996, but since then it has been constantly and substantially improved, especially regarding the manufacturing techniques (see later in section \ref{MM_readout_plane} for more details). The first real application of Micromegas detectors in a Rare Event experiment has been in the CERN Axion Solar Telescope (CAST) at CERN \cite{Zioutas, Andriamonje,Arik}. CAST used a single low-background Micromegas detector since 2002 \cite{Abbon} and three of them since 2007 \cite{Aune_I}. CAST has been a test ground for these detectors, where they have been combined with low background techniques like the use of shielding, radiopurity screening of detector components or advanced event discrimination techniques. The last generation of CAST Micromegas detectors are providing outstanding low levels of background \cite{Galan}. The experience in CAST has clearly proved the interest of Micromegas detectors for low background applications.

This has already been acknowledged by a number of groups that are proposing or considering Micromegas-based TPCs for Rare Event searches. In WIMP dark matter searches, the above mentioned DRIFT collaboration has recently demonstrated experimentally that the experiment's basic technique (the negative ion TPC concept) works with Micromegas planes \cite{Lightfoot}, and could
consider Micromegas for future prototypes. The MIMAC experiment \cite{Santos} has successfully built a micro-TPC with Micromegas readout to measure nuclear recoil tracks in Helium. The Japanese NEWAGE experiment \cite{Nishimura} is developing a similar detector with a micro-dot readout, another type of MPGD. The DMTPC collaboration, although testing a different approach based on optical readouts (CCD), uses also a Micromegas-like structure to produce the scintillation to be read \cite{Dujmic}. In the field of $0\nu\beta\beta$ decay recent proposals like those of NEXT \cite{next} or EXO \cite{Gornea} collaborations have resuscitated the old xenon gas TPC but in the context of the newest advances in TPC readouts, and are considering Micromegas as an option for their TPCs. Very recent developments seem to gather experimental evidence of enhanced energy resolution for Micromegas detectors also at high energies \cite{Dafni}, when compared to MWPC or other MPGDs, which make them particularly suitable for $0\nu\beta\beta$ searches.

\section{Micromegas readout planes} \label{MM_readout_plane}

The Micromegas \cite{Giomataris_II, Giomataris_III} readouts make use of a metallic micromesh suspended over a
(usually pixelised) anode plane by means of isolator pillars, defining an amplification gap of the order of 50 to 150 \textrm{$\mu$m}. Electrons drifting towards the readout, go through the micromesh holes and trigger an avalanche inside the gap, inducing detectable signals both in the anode pixels and in the mesh. The particular Micromegas amplification geometry has numerous advantages that have been studied and widely advertised in the literature. They are specially relevant in the context of Rare Events. For instance, it is known \cite{Giomataris_III} that the way the amplification develops in a Micromegas gap is such that its gain $G$ is less dependent on geometrical factors (the gap size) or environmental ones like the temperature or pressure of the gas than in conventional MWPCs or other MPGDs based on charge amplification. This fact allows in general for higher time
stability and spatial homogeneity in the response of Micromegas, facilitating the scaling-up of prototypes as well as long term
operation, both issues crucial in Rare Event experiments. This same effect is responsible of the generally better energy
resolution obtained with Micromegas planes, also a valuable feature in general for Rare Events, as the detector
usually works as calorimeter. In particular, it is a very important feature for $0\nu\beta\beta$ searches, as only a good energy
resolution of the detectors will allow separation of the peak of the $0\nu\beta\beta$ decay from the tail of the standard
$2\nu\beta\beta$ decay (with emission of two neutrinos), which represent an irreducible background to the searched
$0\nu\beta\beta$ mode \cite{Avignone}.

Practical manufacturing and operation of Micromegas detectors have been extremely facilitated by the development of fabrication processes which yield an all-in-one readout, in opposition to ``classical'' first generation Micromegas, for which the mesh was mechanically mounted on top of the pixelised anode. Nowadays most of the applications of the Micromegas concept for applications in particle, nuclear and astroparticle physics, follow the so-called \textit{bulk}-Micromegas type of fabrication method or the more recent \textit{microbulk}-Micromegas.

The \textit{bulk} Micromegas \cite{Giomataris_II} provides large area, robust readouts by integrating a commercial woven wire mesh together with the anode carrying the strips or pixels. This is achieved by means of a photo resistive film having the right thickness, with which the mesh is laminated at high temperatures together with the anode printed circuit, forming a single object. By photolithographic methods the photo-resistive material is etched removing the material between the mesh and the anode and thus creating the Micromegas gap in all the readout area but in some specific spots which are left as supporting pillars. The result is a robust, homogeneous and reproducible readout that can be made up to relatively large areas and low cost, and very little dead space. The technique has achieved maturity and \textit{bulk}-Micromegas in large quantities are being used in several particle physics experiments, like for example in the T2K experiment \cite{Lux}. Single unit \textit{bulk}-Micromegas of $2 \times 1$ \textrm{m$^{2}$} are being envisaged for the MAMMA project \cite{Polychronakos}.

The \textit{microbulk} Micromegas is a more recent development performed jointly by CEA and CERN \cite{Aune_II}. It provides, like the \textit{bulk}, all-in-one readouts but out of double-clad kapton foils. The mesh is etched out of one of the copper layers of the foil, and the Micromegas gap is created by removing part of the kapton by means of appropriate chemical baths and photolithographic techniques. Although the fabrication technique is still under development, the
resulting readouts have very appealing features, outperforming the \textit{bulk} in several aspects. The mechanical homogeneity of the gap and mesh geometry is superior, and in fact these Micromegas have achieved the best energy resolutions among MPGDs with charge amplification. Because of this, time stability of \textit{microbulk} is expected to be also better than \textit{bulk}. On the other hand, they are less robust than the \textit{bulk} and at the moment the maximum size of single readouts are of only $30 \times 30$ \textrm{cm$^{2}$} (the limitation coming from equipment and not fundamental). This type of readouts is being used in the CAST experiment \cite{Abbon}.

However, the most relevant feature of \textit{microbulk}-Micromegas for the purpose of this work and in general for application in Rare Events is their potential radiopurity. The readout can be made extremely light and most of the raw material is kapton and copper, two materials known to be (or to achieve) the best levels of radiopurity \cite{ILIAS}. Of course, the experience in radiopurity tells that every sample needs to be measured and that materials present in negligible amounts can account, if they have high level of radioactive contamination, for the majority of the radioactivity budget of a sample. It is necessary to stress that the fabrication methods followed up to now have not been optimized from the point of view of radiopurity. Therefore, the results here presented have to be taken as typical values for current Micromegas detectors, expecting to achieve higher radiopurity levels by selecting better raw materials or by improving the fabrication processes. A better understanding of these ways of improvement is one of the goals of the present work.

\section{Detector set-up and measurements} \label{setup}

\begin{figure*}[]
\centering
\includegraphics[width=0.4\textwidth]{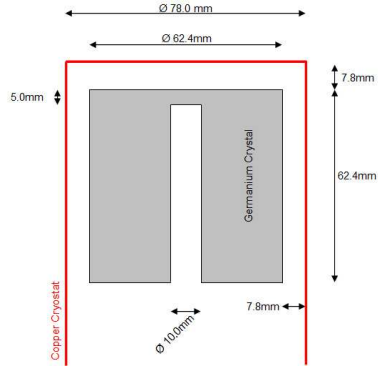}
\caption{Diagram and dimensions of the germanium detector used to carry out the measurements presented in this work and its position inside the copper cryostat.}
\label{Esquema_paquito}
\end{figure*}

\begin{figure*}[t]
\centering
\includegraphics[width=1.\textwidth]{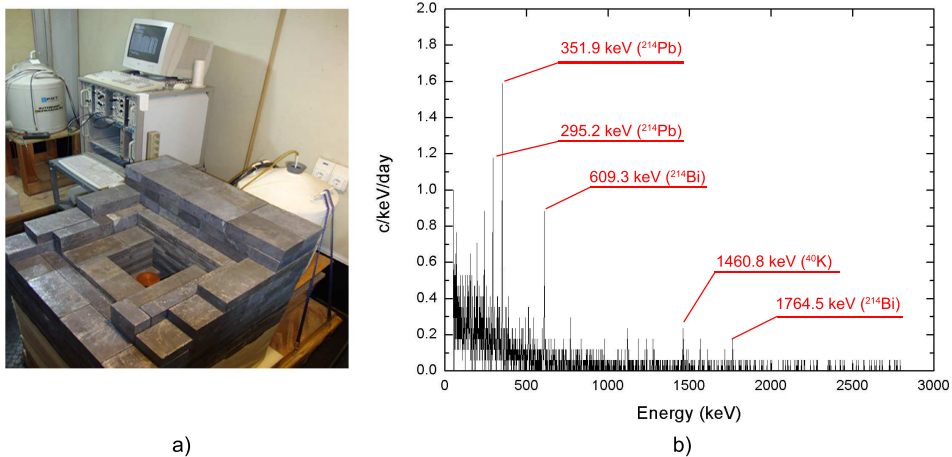}
\caption{a) Picture of the set-up at the Canfranc Underground Laboratory, showing the Ge detector inside the open lead shielding. b) Energy spectrum of a background measurement taken during $\sim$1 month.}
\label{Paquito}
\end{figure*}

All the measurements presented in this work were done using a p-type germanium detector with closed-end coaxial geometry and a volume close to 190 \textrm{cm$^{3}$} and an approximate mass of 1 \textrm{kg}. The crystal is encapsulated in a copper cryostat of 7.8 \textrm{cm}-diameter. Figure \ref{Esquema_paquito} shows a diagram of the detector and its dimensions and the position inside the copper cryostat. The detector is placed inside a shielding composed by 10 \textrm{cm} of archeological lead and 15 \textrm{cm} of low radioactivity lead (Fig. \ref{Paquito}a). The lead box is closed with a PVC bag and a flux of nitrogen gas is continuously introduced to prevent air intrusion. The detector is located at the Canfranc Underground Laboratory (depth 2500 \textrm{meters water equivalent}) to avoid cosmic radiation.

The pulses coming from the detector, with an operation voltage of 3500 \textrm{V}, pass through an electronic chain composed by a Canberra 2001 preamplifier, a linear amplifier Canberra 2020 and an Analog to Digital Converter (ADC) Canberra 8075, that is connected to a PC. The DAQ system of the detector provides information about energy and time of each detected event to allow reconstruction of the energy spectrum, and also to study their time correlation and further remove artificial events, such as electric noise. The detector has an intrinsic resolution (Full Width at Half Maximum) of 2.20 \textrm{keV} at 122 \textrm{keV} and 2.70 \textrm{keV} at 1332 \textrm{keV}. To determine the detection efficiency, measurements with calibrated gamma sources ($^{152}$Eu and $^{40}$K) at different positions have been made; in addition, a GEANT4 simulation application has been developed including specifically the detector set-up. The validity of this application has been checked simulating all the calibrations done with the gamma sources and comparing the corresponding efficiency curves. The agreement between simulated and experimental curves is higher than 95 $\%$ in all the cases. A simulation implementing the measured sample (specially geometry and composition) is carried out to determine the efficiency, considering a 5 $\%$ error deduced from the preliminary tests. Using all the described set-up, a background level of 0.5 \textrm{c/10 keV/day} has been measured from 50 to 3000 \textrm{keV}, the most important contributions being those from $^{214}$Bi, $^{214}$Pb and $^{40}$K decays as it can be seen in Fig. \ref{Paquito}b.

\begin{figure*}
\centering
\includegraphics[width=1.\textwidth]{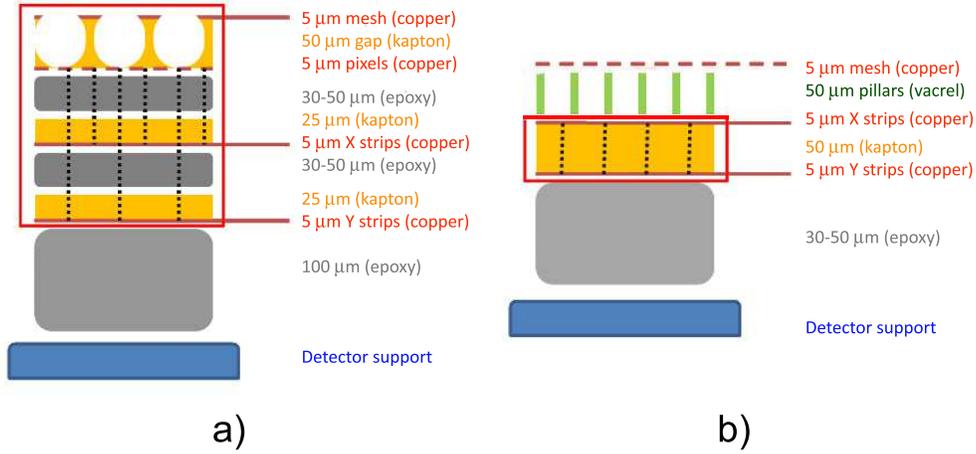}
\caption{Sketch of the two Micromegas from which the samples measured for this study were taken: a) a \textit{microbulk} detector and b) a ``classical'' Micromegas structure. The measured samples include only the components squared in red.}
\label{Esquemas}
\end{figure*}

The samples measured and studied could be divided in two groups. The first of them includes a full \textit{microbulk}-Micromegas readout plane formerly used in the CAST experiment (Fig. \ref{Esquemas}a). The readout has a complex structure made of three kapton layers (and the corresponding intermediate copper ones) glued together in order to constitute the mesh, pillars, anode pixels and necessary underlying strip layout. The scheme of the structure of the different layers can be seen in Fig. \ref{Esquemas}a. The main components of this readout are copper, kapton (double copper clad polyamide laminate Sheldahl G2300) and epoxy (Isola DE156) and it has to be noted that the readout has been removed from its support, implying that some additional traces of epoxy, used to glue the readout to its support, could remain on it. The second sample of this group is part of a classical Micromegas structure. Here, there is only the anode layer of a readout that has been used in the first phase of the CAST experiment, and there is no mesh in the readout (it has to be noted that in this type of classical Micromegas, the mesh was mechanically attached on top of the anode plane as shown in Fig. \ref{Esquemas}b). For this sample, the main materials are copper and kapton, although again there could be some traces of epoxy since the detector had also been glued on a support. For the final purpose of this work, this second sample represents an earlier stage in the manufacturing process than the full \textit{microbulk} structure of the first sample, in which further chemical baths have been applied to etch the kapton pillars and the mesh structure. Both samples of this group have a diameter of 11 \textrm{cm} due to their former application at the CAST experiment \cite{Abbon}, while the thicknesses are $\sim$145 \textrm{$\mu$m} for the \textit{microbulk} detector and $\sim$60 \textrm{$\mu$m} for the classical Micromegas structure (see Fig. \ref{Esquemas}). The masses are 2.8 \textrm{gr} for the full \textit{microbulk}-Micromegas and 2.6 \textrm{gr} for the classical one.

The second group of samples represents the raw material used in the fabrication of \textit{microbulk} readouts. Both samples consist of kapton metalised with copper on one or both sides. One of them has a 25-\textrm{$\mu$m}-thick kapton layer together with a 5-\textrm{$\mu$m}-thick copper one (kapton-copper sample), while the second one combines two 5-\textrm{$\mu$m}-thick copper layers with a 50-\textrm{$\mu$m}-thick  kapton one (copper-kapton-copper sample). In order to have samples as similar as possible to the readouts of the first group, four circles of material of the same diameter (11 \textrm{cm}) have been taken, with corresponding masses of 5.9 \textrm{gr} for the kapton-copper sample, and 4.7 \textrm{gr} for the copper-kapton-copper one. The main motivation to consider this group of samples is to determine if the contamination present in a detector comes from the raw materials (kapton and copper mainly) or from the different treatments needed to complete the manufacture of the detectors.

\section{Results and discussion} \label{RandD}

The radioactivity of all the samples described in Section \ref{setup} was measured by placing them close to the closed part of the germanium detector and inside the shielding. The main gamma emission lines present in the spectra taken with  the samples are compared to their corresponding lines in a \textit{background} spectrum (i.e. a spectrum taken without any sample inside the shielding like the one shown in Fig. \ref{Paquito}b) to obtain values, or upper limits, of the activities of the different isotopes. The gamma lines that, in principle, are studied to estimate the activity of each isotope are summarized in Table \ref{Emissions}. The analysis of all these lines is especially interesting for the isotopes belonging to the natural radioactive chains. This allows determination of the activity of the chain but also to detect if there is any isotope breaking the equilibrium.

\begin{table}
\begin{center}
\caption{Gamma lines (in \textrm{keV}) that are in principle studied to determine the radioactive contamination of different sotopes.}
\centering
\begin{tabular}{|c|c|cc|}
\hline
\textbf{Chain}&\textbf{Isotope}&\multicolumn{2}{|c|}{\textbf{Energy (\textrm{keV})}}\\
\hline
\hline
$^{238}$U&$^{234}$Th&92.5&\\
&$^{226}$Ra&186.2&\\
&$^{214}$Pb&295.2&351.9\\
&$^{214}$Bi&609.3&1764.5\\
\hline
&$^{235}$U&143.8&185.7\\
\hline
$^{232}$Th&$^{228}$Ac&338.3&911.2\\
&$^{228}$Th&84.4&\\
&$^{212}$Pb&238.6&\\
&$^{208}$Tl&583.2&2614.5\\
\hline
&$^{60}$Co&1173.2&1332.5\\
\hline
&$^{40}$K&1460.8&\\
\hline
\end{tabular}
\label{Emissions}
\end{center}
\end{table}

The required measuring times are strongly related to the contamination of the sample in order to reach a good sensitivity level. For this reason, samples expected to have higher levels of contamination need less measuring time than more radiopure samples to determine the activities. In any case the best scenario is to make measurements long enough, so that the background level of the detector and its intrinsic sensitivity is the limiting factor of the radioactivity levels obtained, independently of the measured sample. Taken this property into account, \textit{microbulk}-Micromegas and Micromegas without mesh were measured during 34 and 32 days respectively, the kapton-copper layer during 34 days and the copper-kapton-copper layer during 37 days, while the background spectra used for the analysis correspond to around 30 days of measurement taken before and after the sample measurements. The obtained results, expressed in activity levels per unit surface, are presented for the radioactive chains ($^{232}$Th, $^{235}$U, $^{238}$U), and for $^{40}$K and $^{60}$Co isotopes, being summarized in Table \ref{Results}. It has been checked that the same activity is obtained from all the gamma lines analyzed corresponding to different isotopes of the same radioactive chain (summarized in Table \ref{Emissions}). This is an indication that that $^{232}$Th, $^{235}$U and $^{238}$U chains are in equilibrium.

\begin{table}[]
\begin{center}
\caption{Radioactivity levels (in \textrm{$\mu$Bq/cm$^{2}$}) measured for a Micromegas without mesh, a \textit{microbulk}-Micromegas, a kapton-copper raw material foil, a copper-kapton-copper raw material foil and those in a PMT used in XENON experiment, taken from \cite{Aprile}.}
\label{Results}
\resizebox*{1.\textwidth}{!}{
\centering
\begin{tabular}{|c||c|c|c|c|c|}
\hline
\textbf{Sample}&\textbf{$^{232}$Th}&\textbf{$^{235}$U}&\textbf{$^{238}$U}&\textbf{$^{40}$K}&\textbf{$^{60}$Co}\\
\hline
\hline
Micromegas&&&&&\\
without mesh&4.6$\pm$1.6&$<$6.2&$<$40.3&$<$46.5&$<$3.1$^{*}$\\
\hline
\textit{Microbulk}-Micromegas&$<$9.3&$<$13.9&26.3$\pm$13.9&57.3$\pm$24.8&$<$3.1$^{*}$\\
\hline
kapton-copper foil&$<$4.6$^{*}$&$<$3.1$^{*}$&$<$10.8&$<$7.7$^{*}$&$<$1.6$^{*}$\\
\hline
copper-kapton-copper foil&$<$4.6$^{*}$&$<$3.1$^{*}$&$<$10.8&$<$7.7$^{*}$&$<$1.6$^{*}$\\
\hline
\hline
Hamamatsu&&&&&\\
R8520--06 PMT \cite{Aprile}&27.9$\pm$9.3&-&$<$37.2&1705.0$\pm$310.0&93.0$\pm$15.5\\
\hline
\multicolumn{6}{l}{$^{*}$Level obtained from the Minimum Detectable Activity (MDA) of the detector \cite{Currie}.}\\
\end{tabular}
}
\end{center}
\end{table}

Table \ref{Results_ppb} summarizes the contamination for $^{232}$Th and $^{238}$U chains in ppb and for $^{40}$K in \textrm{ppm} units in order to facilitate  the comparison  with other measurements present in the literature, although it has to be stressed that the samples measured are not homogeneous. For this reason the presented contamination levels could be not necessarily associated to the bulk of the samples, as a \textrm{ppb} quantity induce to consider, and it could be for example a surface contamination.

\begin{table}[]
\begin{center}
\caption{Radioactivity levels (in \textrm{ppb} for $^{232}$Th and $^{238}$U and in \textrm{ppm} for $^{40}$K) measured for a Micromegas without mesh, a \textit{microbulk}-Micromegas, a kapton-copper raw material foil and a copper-kapton-copper raw material foil.}
\label{Results_ppb}
\resizebox*{1.\textwidth}{!}{
\centering
\begin{tabular}{|c||c|c|c|}
\hline
\textbf{Sample}&\textbf{$^{232}$Th}&\textbf{$^{238}$U}&\textbf{$^{40}$K}\\
\hline
\hline
Micromegas&&&\\
without mesh&41.4$\pm$14.4&$<$119.3&$<$54.9\\
\hline
\textit{Microbulk}-Micromegas&$<$72.3&67.3$\pm$35.6&58.5$\pm$25.3\\
\hline
kapton-copper foil&$<$72.9$^{*}$&$<$56.4&$<$16.0$^{*}$\\
\hline
copper-kapton-copper foil&$<$91.5$^{*}$&$<$70.8&$<$20.1$^{*}$\\
\hline
\multicolumn{4}{l}{$^{*}$Level obtained from the Minimum Detectable Activity (MDA) of the detector \cite{Currie}.}\\
\end{tabular}
}
\end{center}
\end{table}

The results do not show radioactive contamination above the minimum detectable level (MDA) of the set-up for the samples of raw material. For the Micromegas samples, some contaminations just above the MDA are seen. These values are in overall compatible between the two samples, although maybe slightly higher in $^{40}$K for the case of the full microbulk readout. This fact leads to the hypothesis that current radioactive control of \textit{microbulk} Micromegas is dominated by the added materials in the composition of the different layers (like epoxy), or by the chemical treatment for the etching processes. This hypothesis should be the starting point of future work to further enhance the radiopurity of these readouts.

Nevertheless, the measured levels of radiopurity are already extremely positive. Although the requirements on radiopurity of a detector component depend on the specific application in question as well as on the expected sensitivity of the experiment, we can generically compare our measurements with typical values of surface radioactivity achieved in other kinds of readout components. Just for illustration, Table \ref{Results} shows radiopurity measurements performed on Photomultiplier tubes (PMTs) which are currently in use or considered in experiments or projects of Rare Event searches, all of them with the highest requirements on radiopurity. In particular, the Hamamatsu PMT R8520--06 is being used by the XENON collaboration in their detectors  \cite{Aprile}, and has been the outcome of a dedicated development to produce radiopure PMT. An array of these devices are at the core of the XENON10 detector, currently the most sensitive dark matter experiment, composing its light readout. They are also being used in the bigger, and more sensitive, XENON100 detector, currently taking data at Gran Sasso. The measured radioactivity per unit surface of Micromegas readouts is substantially lower that those reported for the R8520--06 PMT, especially for $^{40}$K and $^{60}$Co. Although they are not compatible (PMTs are developed for light detection and Micromegas are charge readouts), the reference values of the R8520--06 PMT radiopurity are illustrative as the very low levels of surface radiopurity that are tolerable for leading low background experiments. Therefore the state-of-the-art \textit{microbulk} Micromegas are therefore very well qualified for an ultra low background experiment.

The nature of the selected samples, chosen without any special criterion regarding manufacture or radiopurity, leads to reasonably expect that even lower radioactivity levels could be achieved after some development work to identify and eventually remove (or substitute) the limiting components or treatments. Therefore, the planned work for the near future includes further measurements of individual materials (glues, chemical baths, ...) used in the manufacture of the planes from the raw materials, to identify the origin of the remaining contamination and, if possible, the identification and use of cleaner alternatives. Since the contamination levels obtained from the samples of raw materials (composed by kapton and copper) are limited by the sensitivity of the measurements, the use of alternative measurement techniques like mass spectrometry, neutron activation or improved Ge detectors are contemplated.

\section{Conclusions} \label{Conclusions}

Micromegas readouts are being considered or used in several Rare Event searches, like detection of Dark Matter or $0\nu\beta\beta$, due to important features like its robustness, stability and improved energy resolution. Since one of the most important requirements for a detector focused in Rare Event searches is a high level of radiopurity, it is important to know, as much as possible, the quantity, nature and origin of the radioactive contamination of the readouts.

With this aim, two different samples of Micromegas readouts have been screened, together with some raw materials used for their manufacture, using an ultralow background germanium detector located at LSC. The obtained results have revealed very low levels of radioactivity coming from natural chains ($<$9.3 $\mu$Bq/cm$^{2}$ of $^{232}$Th,$<$13.9 $\mu$Bq/cm$^{2}$ $^{235}$U and 26.3$\pm$13.9 $\mu$Bq/cm$^{2}$ $^{238}$U) being $^{40}$K the highest contamination (57.3$\pm$24.8 $\mu$Bq/cm$^{2}$). These levels of radiopurity, compatible for both readouts measured, already fulfil the requirements demanded by the current most stringent low background experiments, demonstrating the high potentiality of the state-of-the-art of Micromegas in the search for Rare Events. The fact that the contamination levels of the raw materials are below the sensitivity of the detector used, indicates that the main contribution to the radioactive contamination (specially $^{40}$K) comes from elements used for the treatment of the readouts and not from the raw materials. Future work will focus on measurements of the component materials separately trying to enclose the main radioactivity sources in the readout and to identify possible processes which add impurities, even using complementary measuring techniques like mass spectrometry or neutron activation analysis to obtain more sensitive results.

\section{Acknowledgments} \label{Acknowledgements}

We want to thank our colleagues of the group of the University of Zaragoza, as well as to colleagues from the CAST, NEXT and RD-51 collaborations for helpful discussions and encouragement. We acknowledge support from the Spanish Ministry of Science and Innovation (MICINN) under contract ref. FPA2008-03456, as well as under the CUP project ref. CSD2008-00037 and the CPAN project ref. CSD2007-00042 from the Consolider-Ingenio2010 program of the MICINN. We also acknowledge support from the European Commission under the European Research Council T-REX Starting Grant ref. ERC-2009-StG-240054 of the IDEAS program of the 7th EU Framework Program. Finally, we also acknowledge support from the Regional Government of Arag\'{o}n under contract PI001/08.



\begin{thebibliography}{100}

\bibitem{Irastorza}
  I. G. Irastorza, Proceedings of XXXVII International Meeting on Fundamental Physics (IMFP2009), 9th-13th February 2009, Benasque (Spain). arXiv0911.2855.

\bibitem{Avignone}
  F. T. Avignone III {\it et al.}, New J. of Phys. {\bf 7} (2005) 6.

\bibitem{Heusser}
  G. Heusser {\it et al}, Annu. Rev. Nucl. Part. Sci. {\bf 45} (1995) 543--590.

\bibitem{Formaggio}
  J. A. Formaggio {\it et al}, Annu. Rev. Nucl. Part. Sci. {\bf 54} (2004) 361--412.

\bibitem{Ahlen}
  S. Ahlen {\it et al}, Int. J. of Mod. Phys. A {\bf 25} (2010) 1--51.

\bibitem{Spergel}
  D. N. Spergel, Phys. Rev. D {\bf 37} vol. 6 (1988) 1353--1355.

\bibitem{Pipe}
  S. Burgos {\it et al}, Astropart. Phys. {\bf 31} (2009) 261--266.

\bibitem{Dujmic}
  D. Dujmic {\it et al}, Astropart. Phys. {\bf 30} (2008) 58--64.

\bibitem{Luscher}
  R. Luscher {\it et al}, Phys. Lett. B. {\bf 434} (1998) 407--414.

\bibitem{Oed}
  A. Oed, Nucl. Instr. and Meth. A {\bf 263} (1988) 351--359.

\bibitem{Giomataris_I}
  I. Giomataris {\it et al}, Nucl. Instr. and Meth. A {\bf 376} (1996) 29--35.

\bibitem{Zioutas}
  K. Zioutas {\it et al.} [CAST Collaboration], Phys. Rev. Lett. {\bf 94} (2005) 121301.

\bibitem{Andriamonje}
  S. Andriamonje {\it et al.}  [CAST Collaboration], JCAP {\bf 0704} (2007) 010.

\bibitem{Arik}
  E. Arik {\it et al.}  [CAST Collaboration], JCAP {\bf 0902} (2009) 008.

\bibitem{Abbon}
  P. Abbon {\it et al.}, New J. of Phys. {\bf 9} (2007) 170.

\bibitem{Aune_I}
  S. Aune {\it et al.}, Nucl. Instr. and Meth. A {\bf 604} (2009) 15--19.

\bibitem{Galan}
  J. Gal\'{a}n {\it et al.}, JINST {\bf 5} (2010) P01009.

\bibitem{Lightfoot}
  P. K. Lightfoot, {\it et al.}, Astropart. Phys. {\bf 27} (2007) 490--499.

\bibitem{Santos}
  D. Mayet {\it et al.}, J. of Phys.: Conf. Ser. {\bf 65} (2007) 012012.

\bibitem{Nishimura}
  K. Miuchi {\it et al.}, Phys. Lett. B. {\bf 686} (2010) 11--17.

\bibitem{next}
  Letter of Intent to the Canfranc Underground Laboratory, NEXT collaboration, arXiv:0907.4054
[hep-ex].

\bibitem{Gornea}
  R. Gornea {\it et al.}, J. of Phys.: Conf. Ser. {\bf 179} (2009) 012004.

\bibitem{Dafni}
  T. Dafni {\it et al}, Nucl. Instr. and Meth. A {\bf 608} (2009) 259--266.

\bibitem{Giomataris_II}
  I. Giomataris {\it et al.}, Nucl. Instr. and Meth. A {\bf 560} (2006) 405--408.

\bibitem{Giomataris_III}
  I. Giomataris {\it et al.}, Nucl. Instr. and Meth. A {\bf 419} (1998) 239--250.

\bibitem{Lux}
  T. Lux {\it et al.}, J. of Phys.: Conf. Ser. {\bf 65} (2007) 012018.

\bibitem{Polychronakos}
  V. Polychronakos {\it et al.}, ATL-M-MN-0001 (2007).

\bibitem{Aune_II}
  S. Aune {\it et al.}, J. of Phys.: Conf. Ser. {\bf 179} (2009) 012015.

\bibitem{ILIAS}
  ILIAS database on radiopurity of materials. http://radiopurity.in2p3.fr (2008).

\bibitem{Aprile}
  E. Aprile, J. of Phys.: Conf. Ser. {\bf 203} (2010) 012005.

\bibitem{Currie}
    L. A. Currie, Anal. Chem. {\bf 40} vol. 3 (1968) 586--593.


\end{thebibliography}
\end{document}